\documentclass[preprintnumbers,nofootinbib,floatfix,notitlepage]{revtex4-1}

\usepackage{graphicx}
\usepackage{hyperref}

\begin{document}

\pagestyle{plain}


\title{The Case for Muon-based Neutrino Beams}

\author{Patrick Huber\footnote{email: pahuber@vt.edu}
}
\affiliation{Center for Neutrino Physics, Department of Physics,
  Virginia Tech, Blacksburg, VA 24061, USA}
\author{Alan Bross\footnote{email: bross@fnal.gov}}
\author{Mark Palmer\footnote{email: mapalmer@fnal.gov}}
\affiliation{Fermi National Accelerator Laboratory, Batavia, IL 60510, USA}


\preprint{FERMILAB-FN-0991-APC}


\begin{abstract}

\end{abstract}
\date{\today}

\maketitle

\section{Introduction}
\label{sec:intro}

For the foreseeable future, high energy physics accelerator capabilities in the US will be deployed 
to study the physics of the neutrino sector.  This thrust for the US domestic program was confirmed 
by the recent Particle Physics Project Prioritization Panel (P5) 
report~\cite{ParticlePhysicsProjectPrioritizationPanel(P5):2014pwa}.   In this context, it is useful to 
explore the sensitivities and limiting systematic effects of the planned neutrino oscillation program, 
so that we can evaluate the issues that must be addressed in order to ensure the success of these 
efforts.  It is only in this way that we will ultimately be able to elucidate the fundamental physics 
processes involved.  We conclude that success can only be guaranteed by, at some point in the 
future, being able to deploy muon accelerator capabilities.  Such capabilities provide the only route 
to precision neutrino beams with which to study and mitigate, at the sub-percent level, the limiting 
systematic issues of future oscillation measurements.  Thus this analysis argues strongly for 
maintaining a viable accelerator research program towards future muon accelerator capabilities.

\section{Short-baseline physics}
\label{sec:sbl}

Most models of neutrino mass generation involve right-handed, and
hence sterile, neutrinos, however, the mass of those right-handed
neutrinos is not well constrained. In principle, the mass scale of the
right-handed neutrino can range from sub-eV up to the Planck scale and
only a few regions of parameter space have been probed in
detail. Thus, the existence of sterile neutrinos with a mass around the
eV-scale seems plausible. Experimental results from
LSND~\cite{Athanassopoulos:1996jb} indicate a
$\bar\nu_\mu\rightarrow\bar\nu_e$ flavor conversion at the level of
about $0.003$. Given the baseline and mean energy of neutrinos in this
experiment, oscillation involving a neutrino with a mass squared
difference $\Delta m^2\sim 1\,\mathrm{eV}^2$ is a possible and
straightforward explanation. More recently, the MiniBooNE
experiment~\cite{AguilarArevalo:2008rc,AguilarArevalo:2010wv} has seen
hints of flavor transitions in both $\bar\nu_\mu\rightarrow\bar\nu_e$
and $\nu_\mu\rightarrow\nu_e$, which appear to be consistent with the
oscillation interpretation of the LSND result.

Atmospheric neutrino oscillations imply  $\Delta m^2_{31}\simeq
2\times10^{-3}\,\mathrm{eV}^2$ and solar neutrino oscillations require
$\Delta m^2_{21}\simeq 7\times10^{-5}\,\mathrm{eV}^2$. Therefore, at
least four neutrinos are required to allow for another $\Delta m^2$ to
be of order eV$^2$. From the invisible decay-width of the $Z$-boson, it
is known that there are only 3 active, light neutrinos and hence, the
extra neutrino to explain LSND and MiniBooNE has to be sterile.

Any oscillation from one active neutrino into another active neutrino
mediated by a sterile neutrino requires that the sterile neutrino
mixes with both the initial and final active neutrino flavor. As a
consequence, any appearance signal, as potentially observed by LSND
and MiniBooNE, implies the existence of a corresponding disappearance
signal. This correspondence can be made quantitative: the energy
averaged oscillation probabilities obey the following inequality,
irrespective of the number of sterile neutrinos,
\begin{equation}
\label{eq:sterile}
\langle P_{\nu_\mu\rightarrow\nu_e}\rangle\leq 4 \left(1-\langle P_{\nu_\mu\rightarrow\nu_\mu}\rangle\right)\left(1-\langle P_{\nu_e\rightarrow\nu_e}\rangle\right)\,.
\end{equation} 
An analogous expression holds for antineutrinos, noting that the energy
averaged disappearance probabilities for neutrinos and antineutrinos
are equal.

Somewhat more recently, the reactor antineutrino anomaly has been
noted~\cite{Mention:2011rk}, which indicates a 6\% deficit of
$\bar\nu_e$ from nuclear reactors at distances of
10-100\,m. Approximately one half of the effect, that is 3\% of the
deficit, are due to the re-evaluation of reactor antineutrino
fluxes~\cite{Mueller:2011nm}, which has been independently confirmed
by one of the authors~\cite{Huber:2011wv}. The error budget of the
reactor antineutrino flux calculations is a difficult subject in its
own right, due to the poorly understood impact nuclear structure
might have~\cite{Hayes:2013wra} and the fact that a feature,
the so-called {\it 5 MeV bump}, in the measured spectrum of reactor
antineutrinos has recently been found. This feature is not predicted by the
flux calculations~\cite{Mueller:2011nm,Huber:2011wv}; for a summary
on the 5 MeV bump see Ref.~\cite{Dwyer:2014eka}. Taken at face value,
the 6\% neutrino deficit can be interpreted as disappearance of
$\bar\nu_e$ at a level and with a $\Delta m^2$ consistent with the
LSND and MiniBooNE results and their respective interpretation as
sterile neutrino oscillation.

Support for an eV-scale neutrino also comes from the radioactive
source calibrations of the radiochemical gallium solar neutrino
experiments, GALLEX and SAGE. In order to verify the operation of
these experiments, electron capture sources based on either
chromium-51~\cite{Hampel:1997fc,Abdurashitov:1998ne} or
argon-37~\cite{Abdurashitov:2005tb} were used to expose the detectors
to a well known, mono-energetic $\nu_e$ flux. The resulting number of
germanium atoms stayed below the expectation by about 25\% and this
result can be interpreted as the disappearance of $\nu_e$ with
oscillation parameters consistent with the previously mentioned
evidence~\cite{Giunti:2010zu}.

In combination, these indications have led to a renewed interest in the
question of sterile neutrinos at the eV-scale~\cite{Abazajian:2012ys}
and, as a result, there is a plethora of newly proposed experiments. At
the same time, the fact that no $\nu_\mu$ disappearance at the
relevant value of $L/E$ has been observed and many other searches have
produced null results is a source of significant tension in global
fits, see for instance Ref.~\cite{Kopp:2013vaa}. Also, cosmological
observables are sensitive to the presence of a eV-mass sterile
neutrino and while some authors claim considerable tension, other
authors find acceptable compatibility, {\it
  e.g.}~\cite{Archidiacono:2014apa}. It appears that cosmology so far
remains inconclusive for this problem.

To make any progress on the question of a light sterile neutrino new
experiments are necessary and in recognition of this, recommendation
15 of the P5 report states:
\begin{quote} 
Select and perform in the short term a set of small-scale
short-baseline experiments that can conclusively address experimental
hints of physics beyond the three-neutrino paradigm. Some of these
experiments should use liquid argon to advance the technology and
build the international community for LBNF at Fermilab.
\end{quote}
The key word here is ``conclusively'' and the question arises what it
takes to provide a conclusion to the sterile neutrino saga. The P5
report further recommends that nuSTORM be not part of this (or any
other domestic) portfolio. Conclusively testing the sterile neutrino
interpretation will require sharpening the experimental results on
both sides of Eq.~\ref{eq:sterile} by simultaneously pursuing
appearance and disappearance searches in both neutrino and
antineutrino modes.  Table~\ref{tab:exps} lists all possible
oscillations channels and enumerates the experiments which can access
a given channel. ``SBL'' summarizes all possible experiments which can
be performed in a pion decay in-flight beam and includes all
experiments proposed within the Fermilab short-baseline program. Pion
decay in-flight beams contain an intrinsic $\nu_e$ component at the
level of 1\% which is about 3-10 times larger than the expected
signal. It appears doubtful whether precision measurements in the case of
a discovery are possible in this environment. Due to the poorly known
primary beam flux, any credible disappearance search will require a
near and far detector comparison, which in practice is difficult to
achieve at high accuracy because of the different geometric acceptance
of the near and far detector. OscSNS~\cite{Elnimr:2013wfa} is the
proposal for an experiment exploiting neutrinos from pion
decay-at-rest at the Spallation Neutron Source at Oak Ridge National
Laboratory. OscSNS would use the same process to generate and to
detect neutrinos as LSND did, at the same energy and baseline. This
constitutes the most direct test possible of the original LSND result.
Atmospheric neutrinos already provide stringent limits on $\nu_\mu$
and $\bar\nu_\mu$ disappearance and new experiments like low-energy
extensions of IceCube, see for instance~\cite{Esmaili:2012nz}, as well
as the ICAL detector at the Indian Neutrino Observatory are expected
to significantly improve these limits. SOX is the proposal to deploy
radioactive sources under the Borexino
detector~\cite{Borexino:2013xxa}, currently two different types of
sources are considered. One possibility is a 10\,MCi exposure to a
chromium-51 source, which provides a mono-energetic low-energy $\nu_e$
flux detected by elastic $\nu$-e scattering. Another possibility is a
75\,kCi cerium-144 source, which provides a relatively high-energy
beta-spectrum type flux of $\bar\nu_e$, detected by inverse beta
decay. The cerium source is pursued as an entirely European project
and data taking may start by the end of 2015, whereas the
chromium source would profit from U.S. involvement, specifically
irradation of chromium-50 in the High-Flux Reactor (HIFR) at Oak
Ridge. Both sources are too low in energy to allow for appearance
searches so this would constitute a disappearance search in the
electron channel. IsoDAR~\cite{Bungau:2012ys} also exploits beta
decay as its neutrino source.  In this case it is the high-energy beta
decay of lithium-8, which is produced online using neutrons from a
spallation target driven by a 600\,kW beam of 60\,MeV H$_2^+$
ions. The detection reaction is inverse beta decay and ideally
multi-kiloton detectors based on either water or liquid organic
scintillator are used. All source experiments provide a well-characterized 
source flux and spectrum\footnote{Perhaps with the 
exception of the cerium-144 source, where a few percent of emitted 
neutrinos  stem from forbidden beta decay branches and hence are 
subject to considerable nuclear structure effects.}, however the
energy of the neutrino is entirely determined by nuclear physics. As a result, 
the accessible $L/E$-range is limited. The SOX
configurations suffer from somewhat limited statistics. For IsoDAR,
the question of cost looms large since a significant accelerator
component is required which, in conjunction with the necessary
shielding and decommissioning at the end of the experiment, will
require a careful assessment of the full lifetime
cost. PROSPECT~\cite{Ashenfelter:2013oaa} is one of many proposed
reactor short-baseline experiments aimed at directly confronting the
reactor antineutrino anomaly. The key is to use the near/far detector
concept employed very successfully by Daya Bay and RENO at a much
shorter distance, meters instead of 100's of meters. The resulting
challenge lies in dealing with backgrounds from reactor operation and
surface deployment. Statistical errors can be made quite small thanks
to the enormous flux of reactor antineutrinos, however the $L/E$-range
is limited by the reactor antineutrino spectrum.

\begin{table}[b]
\begin{tabular}{cc|cc}
$\nu_\mu\rightarrow\nu_\mu$&atmospheric, SBL&$\nu_\mu\rightarrow\nu_e$&SBL\\
$\bar\nu_\mu\rightarrow\bar\nu_\mu$&atmospheric, SBL&$\bar\nu_\mu\rightarrow\bar\nu_e$&SBL, OscSNS\\
$\nu_e\rightarrow\nu_e$&SOX&$\nu_e\rightarrow\nu_\mu$&?\\
$\bar\nu_e\rightarrow\bar\nu_e$&PROSPECT, isoDAR, SOX&$\bar\nu_e\rightarrow\bar\nu_\mu$&?\\
\end{tabular}
\caption{\label{tab:exps} List of possible oscillation channels and
  experiments which can access each channel. ``SBL'' refers to all
  short-baseline pion decay in-flight neutrino beams.}
\end{table}
It is quite plausible that a combination of a number of these
experiments will be able to provide a conclusive test of the sterile
neutrino interpretation of the anomalies listed at the beginning. Most
of these experiments will have significant difficulties in going beyond a
simple yes or no answer.  In particular precision studies aimed at
discerning the number of sterile neutrinos involved or studying the
question of potential CP violation for 2 or more sterile neutrinos are
beyond their reach. However, it also is plausible that the sum total
of these experiments would constitute a large investment at a level
similar to nuSTORM, but, as we will argue in the
following paragraph, with significantly less overall capability.

\begin{figure}
\includegraphics[width=0.6\textwidth]{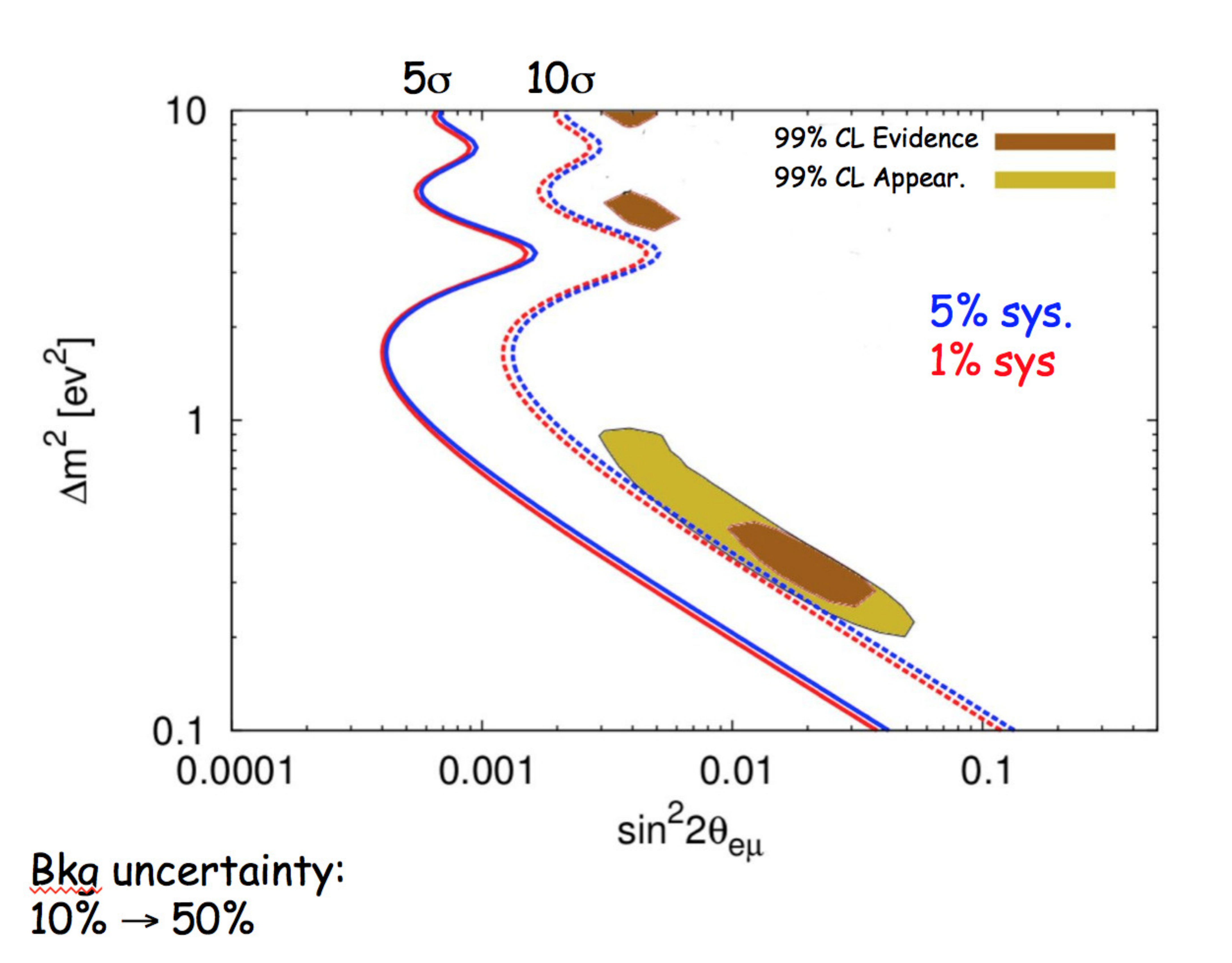}
\caption{\label{fig:nustormsens} Sensitivity of nuSTORM to the
  $\nu_e\rightarrow\nu_\mu$ appearance oscillation due to the presence
  of sterile neutrinos assuming a (3+1) model with anticipated and
  inflated systematics, compared to 99\% confidence contours from
  global fits to the evidence for sterile neutrinos and to all
  available appearance experiments generated by Kopp {\it
    et. al.}~\cite{Kopp:2013vaa} (filled contours) and limits set by
  ICARUS~\cite{Antonello:2013gut}.  Figure and caption adapted from
  Ref.~\cite{Adey:2014rfv}}
\end{figure}

The nuSTORM experiment would provide simultaneous access to all of the
oscillation channels listed in Tab.~\ref{tab:exps}, a unique feature 
among the proposed facilities; for a detailed description see
Ref.~\cite{Adey:2014rfv}. The absolute beam normalization and spectrum
will be known to better than 1\% based on beam instrumentation. Storing
either $\mu^+$ or $\mu^-$ allows for the production of precisely
controlled CP-conjugate beams. Combined with the right suite of
detectors, appearance and disappearance measurements at an
unprecedented and unrivaled accuracy become possible, potentially 
including a very precise neutral current disappearance search. The
sensitivity in the golden appearance mode $\nu_e\rightarrow\nu_\mu$,
the CPT conjugate of the original LSND signal, is shown in
Fig.~\ref{fig:nustormsens}.%
Clearly, a test at very high confidence level is possible. What is
more important is having both sufficiently large statistics and
sufficiently small systematics to explore any possible signal in great
detail, ultimately pinning down the underlying physics.

\section{Long-baseline physics}
\label{sec:lbl}

\begin{figure}
\includegraphics[width=0.6\textwidth]{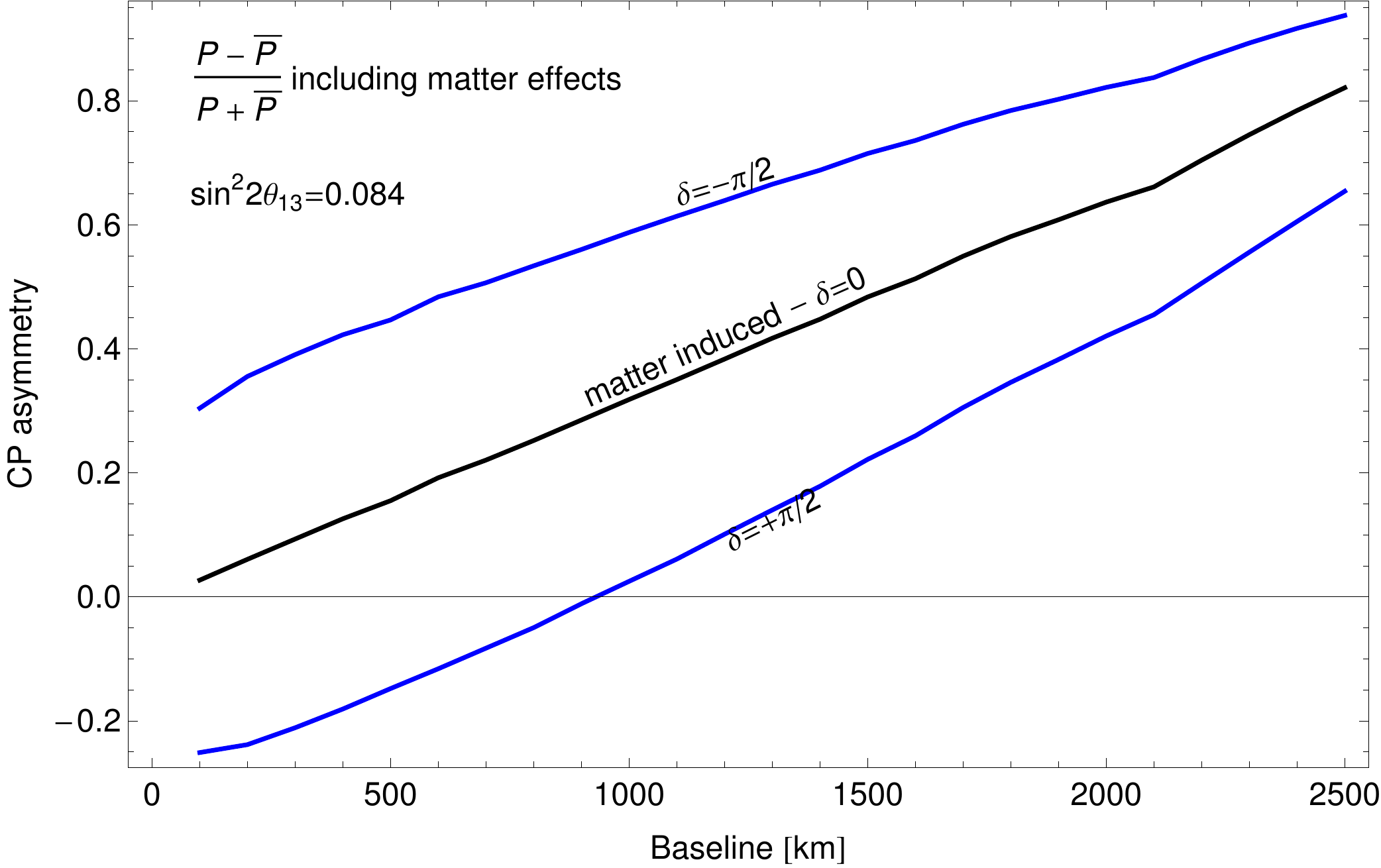}
\caption{\label{fig:cpa}Value of the CP asymmetry $A$ for different
  choices of $\delta$ as a function of the baseline.}
\end{figure}

With the start of the LHC in 2010 and the shutdown of the Tevatron in
2011, the focus of the domestic experimental high-energy physics
program has shifted from the Energy Frontier to the Intensity
Frontier. In order to ensure a vital program, DOE is committed to
making a very significant investment (at the level of a billion
dollars) in new experiments at the Intensity Frontier over the next
decade.  This plan has recently been endorsed by the report of the P5
sub-panel of HEPAP. Specifically, recommendation 13 of the P5 report
reads:
\begin{quote}
Form a new international collaboration to design and execute a highly
capable Long-Baseline Neutrino Facility (LBNF) hosted by the U.S. To
proceed, a project plan and identified resources must exist to meet
the minimum requirements in the text. LBNF is the highest-priority
large project in its timeframe.
\end{quote}
The P5 report further stipulates that the minimum requirement is to
have a sensitivity to discover CP violation for at least 75\% of all
CP phases at the 3\,$\sigma$ confidence level. This requirement translates
directly into an upper limit on the acceptable systematic uncertainty. The
CP asymmetry, $A$, is defined as
\begin{equation}
\label{eq:cpa}
A=\frac{\langle P\rangle-\langle \bar P\rangle}{\langle P\rangle+\langle \bar P\rangle}\,,
\end{equation}
where $\langle P\rangle$ is the energy averaged oscillation
probability for $\nu_\mu\rightarrow\nu_e$ and $\bar P$ is the
corresponding quantity for antineutrinos. The energy average is taken
over the range defined by having one half of the peak probability
around the first oscillation maximum. A rough approximation of the
problem at hand is provided by stating that a measurement of the CP
phase is equivalent to a measurement of the asymmetry $A$, since
$A\propto \sin\delta$, with $\delta$ being the CP phase. For
$\delta\sim 0$ or $\pi$ the error on $\sin\delta$ and $\delta$ are
approximately the same. In Fig.~\ref{fig:cpa} the value of the
asymmetry for different choices of $\delta$ is shown as a function of
the baseline.
Note that, even for a CP-conserving value of $\delta=0$, there is a
non-vanishing asymmetry due to matter effects (this figure assumes
a normal hierarchy).  For baselines below 1500\,km, the genuine CP
asymmetry is at most $\pm 25\%$, whereas for 75\% of the parameter
space in $\delta$, the genuine CP asymmetry can be as small as $\pm
5\%$. That is, a 3$\sigma$ evidence for CP violation in 75\% of
parameter space requires a $\sim 1.5\%$ measurement of the $P-\bar P$
difference. Assuming that the statistical and systematic error
contribute at the same level, a 1\% systematic error is required.

Of course, experiments neither directly measure $\langle P\rangle$
nor the probability $P$.  Instead they measure event rate distributions, $R$, as a
function of the visible energy:
\begin{equation}
\label{eq:rate}
R^\alpha_\beta(E_\mathrm{vis})=N\int dE\,\Phi_\alpha(E)\,\sigma_\beta(E,E_\mathrm{vis})\,\epsilon_\beta(E)\,P(\nu_\alpha\rightarrow\nu_\beta,E) \,,
\end{equation}
where $\alpha$ is the initial neutrino flavor and $\beta$ the
corresponding final flavor. Furthermore, $N$ is the overall
normalization (fiducial mass), $\Phi_\alpha$ is the flux at the
detector of $\nu_\alpha$, $\sigma_\beta$ is the cross section for
$\nu_\beta$, and $\epsilon_\beta$ is the detection efficiency for
$\nu_\beta$. Note that $\sigma_\beta \epsilon_\beta$ always appear in
that combination, hence we can define an effective cross section
$\tilde{\sigma}_\beta:=\sigma_\beta \epsilon_\beta$. $\sigma_\beta$
depends on both the true neutrino energy $E$ and the visible energy
$E_\mathrm{vis}$, since a neutrino of a given energy $E$ can produce a
range of visible energy depending on the underlying event. In practice
the situation is complicated by the detector response which will
translate the visible energy into the reconstructed energy. Even if we
ignore all energy dependencies of efficiencies, cross sections {\it
  etc.}, we generally cannot expect to know any of the fluxes $\phi$
or any of the effective cross sections $\tilde\sigma$ at the required
percent level of accuracy. There generally are no constraints at the
percent level on flux ratios between flavors and/or neutrinos and
antineutrinos. For cross section ratios, it is clear that
neutrino/antineutrino ratios are known only at very high energy
exceeding the 10's of GeV. At low energy, there is no guarantee that
the nuclear cross sections of $\nu_e$ and $\nu_\mu$ scattering are the
same. At the quark level, lepton universality ensures the same coupling
strength at the vertex, but the lepton mass impacts the phase space
and thus the momentum transfer to the nucleus. The response of the
nucleus depends sensitively, and in an essentially not well known
manner, to small changes in $Q^2$. Neutrino experiments cannot measure
the actual $Q^2$ value in an event and thus, effectively only see the
total interaction rates fully integrated over the kinematically
allowed $Q^2$-range. A detailed discussion based on an analysis of
form factors associated with the corresponding hadronic and leptonic
currents has been presented by Day and McFarland~\cite{Day:2012gb}. In their
analysis, differences of several percent in the $\nu_e$ to $\nu_\mu$
cross sections at energies below 1\,GeV appear possible. Furthermore,
even if we may be able to determine $\sigma_e/\sigma_\mu$ from theory, we
will not know the corresponding ratio of efficiencies
$\epsilon_e/\epsilon_\mu$.

The problem that the neutrino flux or cross section is not known with
sufficient accuracy has been encountered many times in neutrino
physics. A proven solution is the use of a near detector to
measure the un-oscillated event rate. In the ratio of far to near
detector data many uncertainties will cancel. In practice, this
requires that near and far detectors are very well understood in their
response and geometrical acceptance. Assuming that the detector response is
identical between near and far detectors and that only total rates matter,
the far/near ratio is given by
\begin{equation}
\label{eq:cancellation}
\frac{R^\alpha_\alpha(\mathrm{far})L^2}{R^\alpha_\alpha(\mathrm{near})}
=\frac{N_\mathrm{far}\Phi_\alpha\,\tilde{\sigma}_\alpha\,
  P(\nu_\alpha\rightarrow\nu_\alpha)}{N_\mathrm{near}
  \Phi_\alpha\,\tilde{\sigma}_\alpha 1 }=\frac{N_\mathrm{far}}{N_\mathrm{near}}\,P(\nu_\alpha\rightarrow\nu_\alpha)\,.
\end{equation}
This method has been applied very successfully in the Daya Bay
experiment to measure $\theta_{13}$~\cite{An:2012eh}, where the
conditions were quite ideal: near and far detectors employ the
same material, size and design, and thus are functionally identical
with sub-percent precision~\cite{DayaBay:2012aa}; near and far
detectors see the reactor cores as point sources; the inverse
beta-decay cross section is independently known; and the initial and final
flavor are the same.

To extrapolate this experience to future beam experiments, the
following factors need to be considered. First, seeing the neutrino
source as point-like will require a not-so-near near detector, which
for baselines of more than a few hundred km requires prohibitively
expensive tunneling. Second, due to the enormous size of the far
detector, the near and far detectors cannot be truly identical. Third, in
a neutrino beam the energy spread is sufficiently large that several
different interaction mechanisms will contribute to the event sample
and thus energy dependencies can be not neglected. For a disappearance
measurement, MINOS provides a good real-world example of the arising
issues and practical methods to deal with
them~\cite{Michael:2006rx}. Even in disappearance mode and employing a
very sophisticated target-horn configuration to systematically cross
check primary neutrino production, the overall systematic error
remains at the level of 1-3\% in neutrino mode~\cite{Adamson:2011ig}
and at a higher level in the antineutrino mode~\cite{Adamson:2012rm}.

For an appearance measurement, the initial and final flavor are
different. Thus, everything else being equal, the cancellation in
Eq.~\ref{eq:cancellation} will be less complete, and a term of the
form $\tilde\sigma_\beta/\tilde\sigma_\alpha$ will remain. Even under
the assumption that the initial flux $\phi_\alpha$ would be well
known, a measurement of $\tilde\sigma_\beta$ in a pure beam of flavor
$\alpha$ is obviously not possible. A realistic pion decay-in-flight
neutrino beam will contain a small admixture of $\nu_e$ but the size
of this admixture is not well constrained across the whole energy range.
Also, the small size of this admixture will limit the statistical accuracy in
the near detector. T2K made an attempt to exploit the $\nu_e$
component in a $\nu_\mu$ beam for a cross section
measurement~\cite{Abe:2014agb}. The total systematic error is about
16\% dominated by the $\nu_e$ beam flux uncertainty and the detector
response. At this point it is unclear how one could 
reduce the systematic error budget to the percent level. The
situation for antineutrinos is likely to be considerably more
difficult. In general, it is far from obvious how to achieve a 1\%
cross section measurement in a beam for which the flux is only known to 5\%.

\begin{figure}
\includegraphics[width=0.5\textwidth]{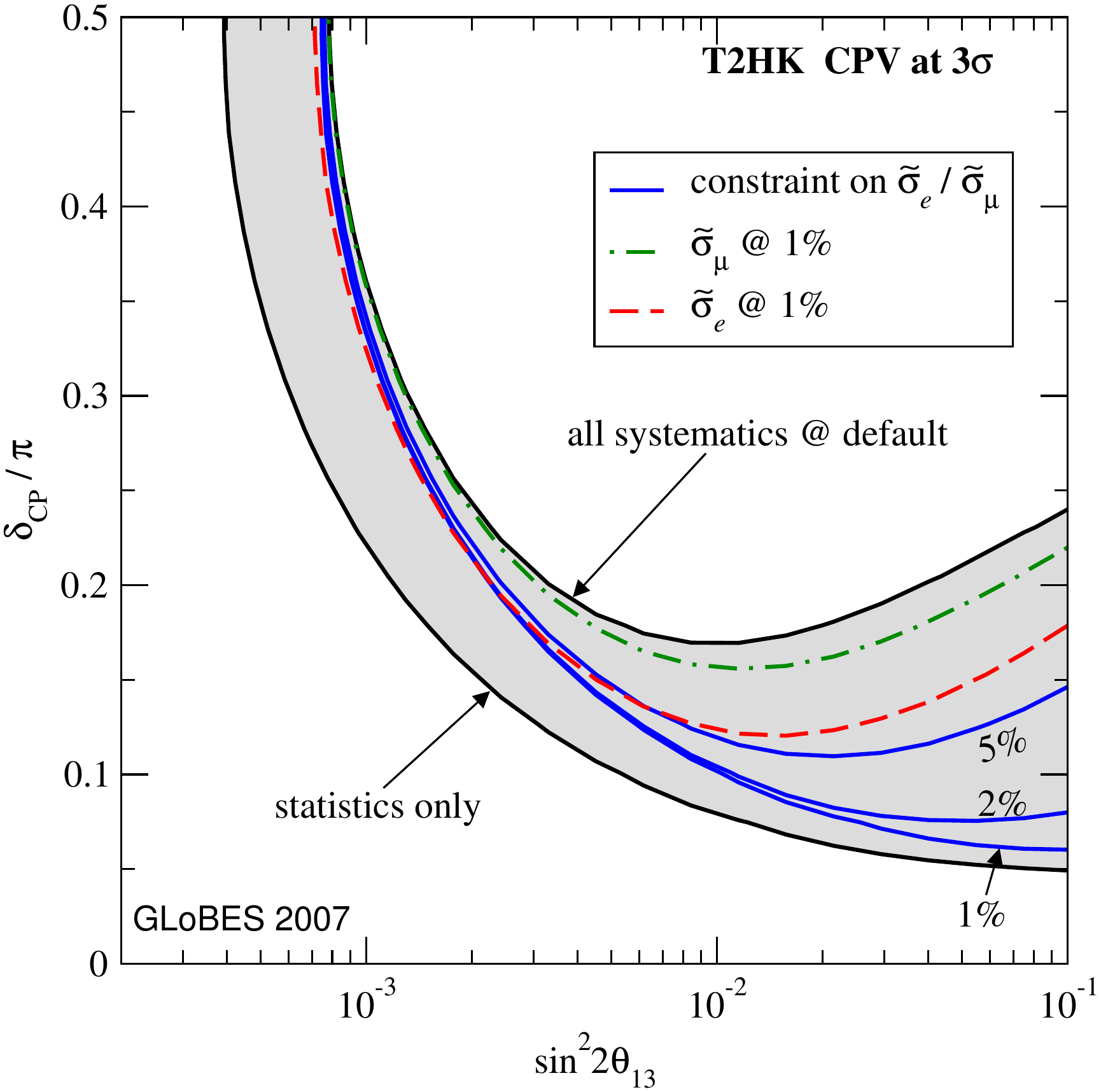}
\caption{\label{fig:t2hk} CP violation sensitivity at 3$\,\sigma$ for
  a certain choice of of systematical errors according and for
  statistical errors only (curves delimiting the shaded region). We
  show also the sensitivity if certain constraints on the product of
  cross sections times efficiencies $\tilde\sigma$ are available: 1\%
  accuracies on $\tilde\sigma_{\mu}$ and $\tilde\sigma_{e}$ for
  neutrinos and antineutrinos, and 5\%, 2\%, 1\% accuracies on the
  ratios $\tilde\sigma_\mu/\tilde\sigma_e$ for neutrinos and
  antineutrinos. Figure and caption adapted from
  Ref.~\cite{Huber:2007em}.}
\end{figure}

To illustrate the quantitative effect on the ability to measure CP
violation, a T2HK-like setup was studied in Ref.~\cite{Huber:2007em},
where a total of more than 20 nuisance parameters, including total
cross section uncertainties, were considered. One of the main results
is shown in Fig.~\ref{fig:t2hk}, where the sensitivity to CP violation
is shown both for statistical errors only as well as for the full systematic
error budget. It is obvious that a constraint on the
$\tilde\sigma_\mu/\tilde\sigma_e$ ratio is the most efficient, and
only, way to recover the desired statistical sensitivity.

Up to this point the whole discussion has been framed in terms of energy
independent quantities, but looking at Eq.~\ref{eq:rate} it is obvious
that the relation between the true neutrino energy $E$ and the visible
energy $E_\mathrm{vis}$ enters prominently into this problem. For the
moment, we will neglect detector effects stemming from finite energy
resolution. Under this simplifying assumption, the problem stems from
the fact that detectors are made of nuclei and not free protons (or
deuterium). This gives rise to a number of significant systematic effects, which
we will summarily refer to as {\it nuclear effects}:
\begin{itemize}
\item Initial state momentum distribution where, due to being in a bound
  state, each nucleon has a non-zero kinetic energy and momentum.
\item Nuclear excitations where, in an interaction between a neutrino and nucleus, some energy
  is transferred into the nuclear system, resulting in, for
  instance, de-excitation gamma rays.
\item Reaction products leaving the nucleus where, when the primary
  vertex is deep inside the nucleus, the reaction products have to
  traverse the nuclear medium resulting in a modified energy
  distribution or complete absorption. It is, for instance, possible
  that a proton is produced but strikes a neutron and the final
  state therefore contains this neutron.
\item Higher order interactions where, given the nuclear bound state,
  there are significant correlations between individual nucleons
  resulting in interactions with more than one nucleon.
\end{itemize}
As a function of $Q^2$ these effects are flavor blind, but $Q^2$ is
\emph{not} measured in a typical long-baseline experiment. As
explained previously, the lepton masses affect the available phase
space and thus the $Q^2$ distribution.  Hence there are flavor effects
in the total cross section. None of these nuclear effects is 
expected to be the same (or even similar) for neutrinos and
antineutrinos. This is exemplified by the very different
$y$-distributions of neutrinos and antineutrino in deep-inelastic
scattering, where $y$ measures the degree of inelasticity. Despite the
fact that the electroweak sector of the Standard Model is extremely
well understood, it is very hard to perform precise computations of
the neutrino-nucleus interactions.  Since existing neutrino beams are
subject to large intrinsic uncertainties, measurements of the neutrino
nucleus cross sections have errors in the 10-30\% range, with
the exception of a few fully inclusive channels. Obtaining a
significantly better quantitative understanding of neutrino cross
sections on nuclear targets is a very hard theoretical problem, since
nuclear structure for large nuclei such as argon is not well
understood. Multi-nucleon correlations as well as final state
interactions have to be correctly included to provide reliable
neutrino cross sections. A number of techniques to derive systematic
approximations exist, but many calculations so far cover only limited portions
of the relevant kinematic regions. A considerable effort is currently
being devoted to the development of theoretical models capable of providing a fully
quantitative description of neutrino-nucleus interactions in the
kinematic regime relevant to the MicroBooNE, LBNE, ArgoNeut and Captain experiments.  
On the other hand, for light nuclei up to carbon, detailed calculations
of the wave function exist, which reproduce measured energy levels
quite accurately. Most of the advances made in our theoretical
understanding of neutrino-nucleus interactions are \emph{not}
available to experiments, since currently used event generators
generally either lag behind theory by decades or are closed,
proprietary codes carefully tuned to the data of the particular
experiment using them.

\begin{figure}
\includegraphics[width=0.5\textwidth]{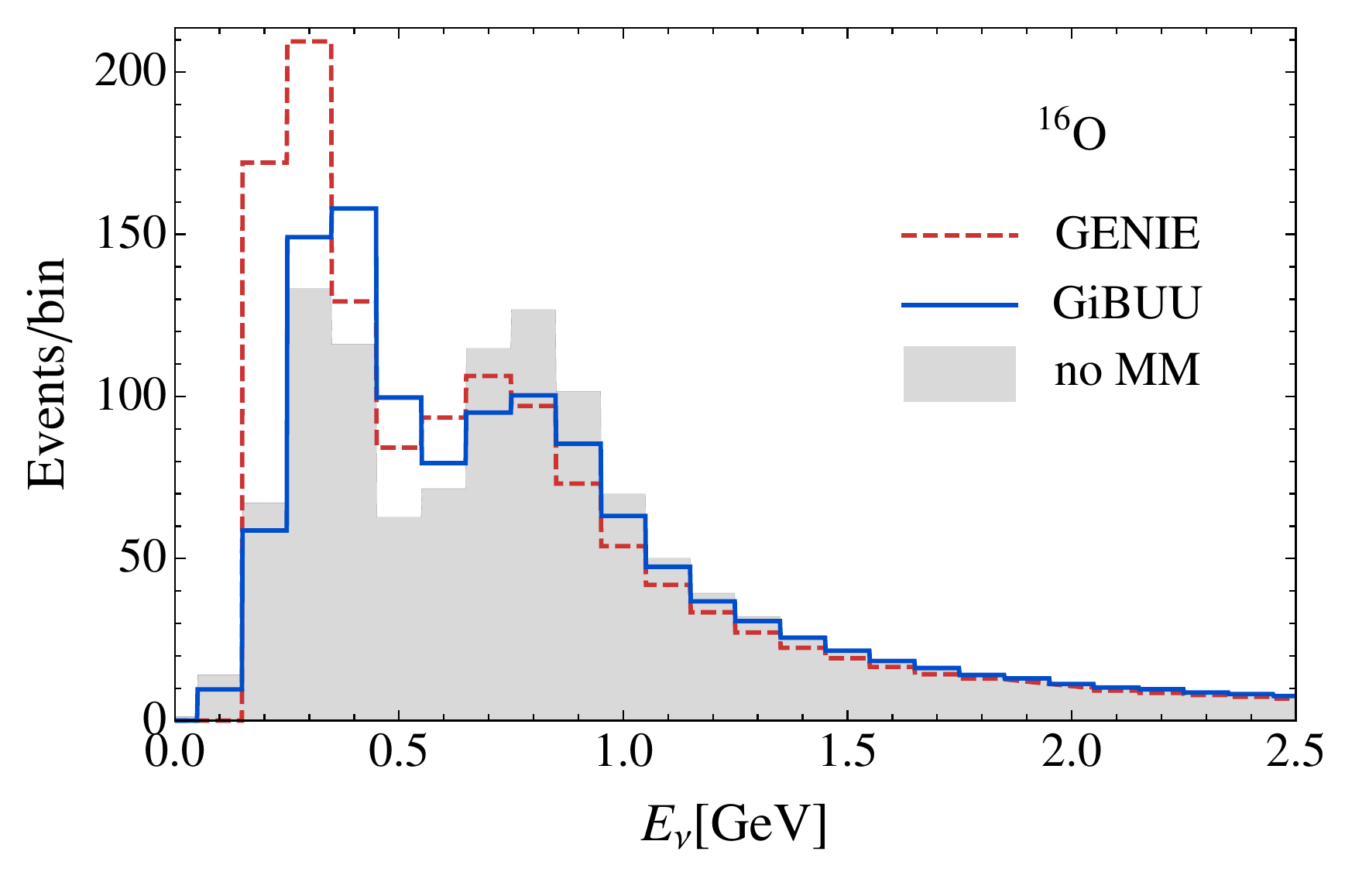}\includegraphics[width=0.5\textwidth]{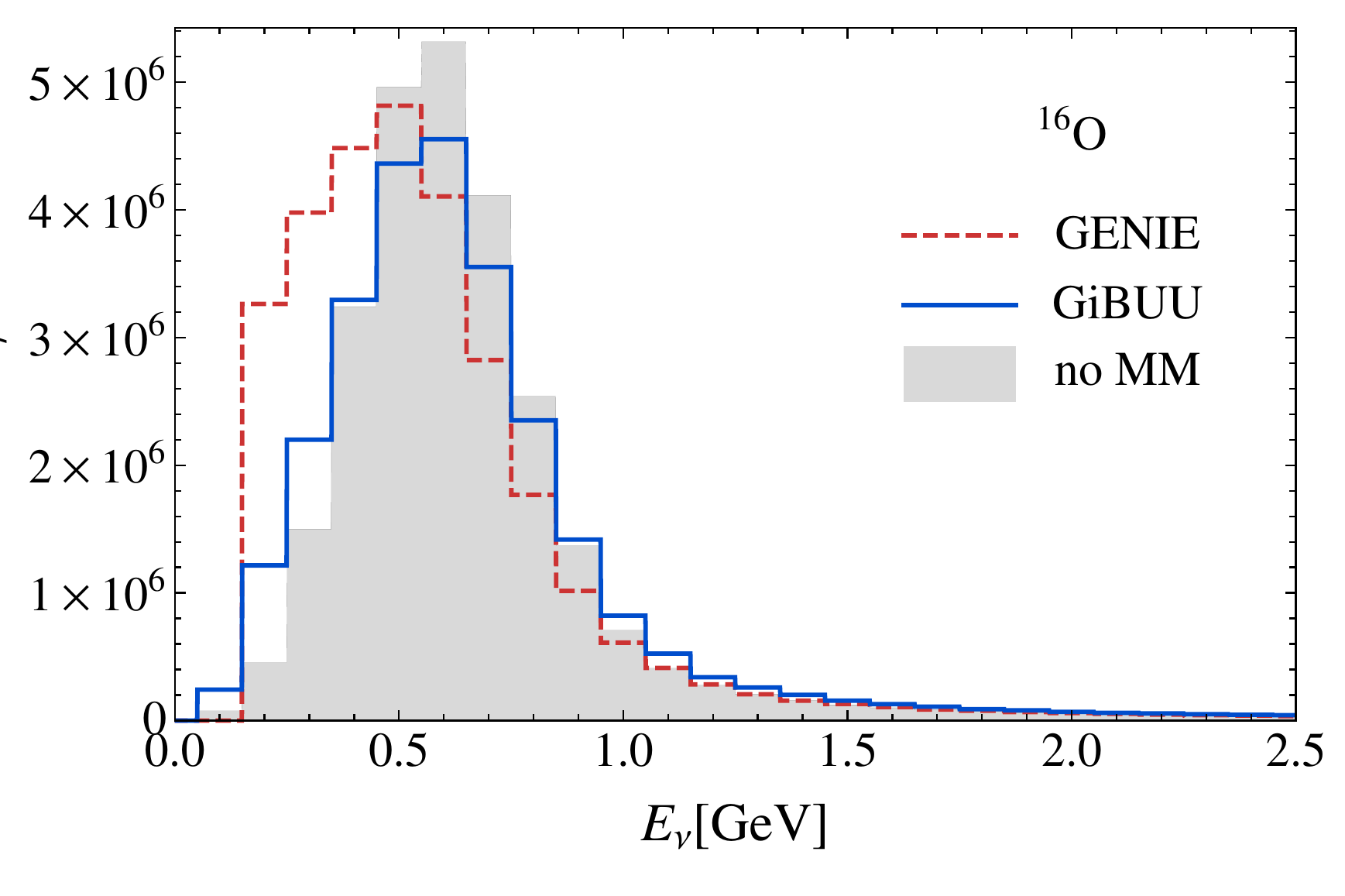}
\caption{\label{fig:qev} Binned QE-like event rates as a function of
  the reconstructed neutrino energy in GeV. The solid blue (dashed
  red) lines show the event rates obtained after migration using the
  GiBUU (GENIE) event generators. The shaded areas show the expected
  event rates coming from the QE-like event sample computed using the
  GiBUU cross-section for $^{16}$O, as for the solid blue lines, but
  without including any migration matrices. For the shaded areas, a
  Gaussian energy resolution function with a constant standard
  deviation of 85\,MeV is added to account for the finite resolution of
  the detector. Left and right panels show the event rates at the near
  and far detectors, respectively. Figure and caption
  adapted from Ref.~\cite{Coloma:2013tba}}
\end{figure}

Only recently have efforts been made to develop a \emph{quantitative}
understanding of nuclear effects and the resulting systematic
uncertainties in the context of current and future long-baseline
experiments, for instance see
Refs.~\cite{Coloma:2013rqa,Coloma:2013tba}. Arguably, the
conceptually simplest example is provided by quasi-elastic
interactions (QE)
\begin{equation}
\nu_l + X \rightarrow X' + p + l^-\,
\end{equation}
where $l$ can be any lepton flavor. Given that $m_X\simeq m_{X'}\gg
E_\nu,E_l$, there is very little energy carried by the recoil and thus
measuring the charged lepton momentum and scattering angle completely
determines the neutrino energy. In the relevant energy range around
1\,GeV, QE cross sections are the best understood theoretically, but,
as we will see, even that understanding is limited. Nuclear effects
will make some non-QE events appear to be QE events. For instance, a
primary vertex may result in the production of a lepton and a pion, where 
the pion subsequently becomes stuck in the nuclear medium and the only
escaping particle is the lepton.  This process results in the \emph{same}
final state as a true QE event. The crucial difference is, of course,
that the simple kinematic relationship between the charged lepton
momentum and scattering angle and the neutrino energy is no longer 
correct. This is illustrated in Fig~\ref{fig:qev}, where the
reconstructed energy distributions of true QE events and those with the same
final state are shown in the near and far detectors of a T2K-like
experiment. For comparison, the event samples of QE-like events are
shown for two different event generators, GiBUU~\cite{Buss:2011mx} and
GENIE~\cite{Andreopoulos:2009rq}. There is an energy offset between
the results from GiBUU and GENIE, which highlights the state of event
generators. The effect of non-QE events is to smear out the
oscillation dip and to move its position.  Both effects directly impact 
the extraction of oscillation parameters at a level comparable to statistical
errors~\cite{Coloma:2013rqa}, even in presence of a near
detector.

\begin{figure}
\includegraphics[width=0.5\textwidth]{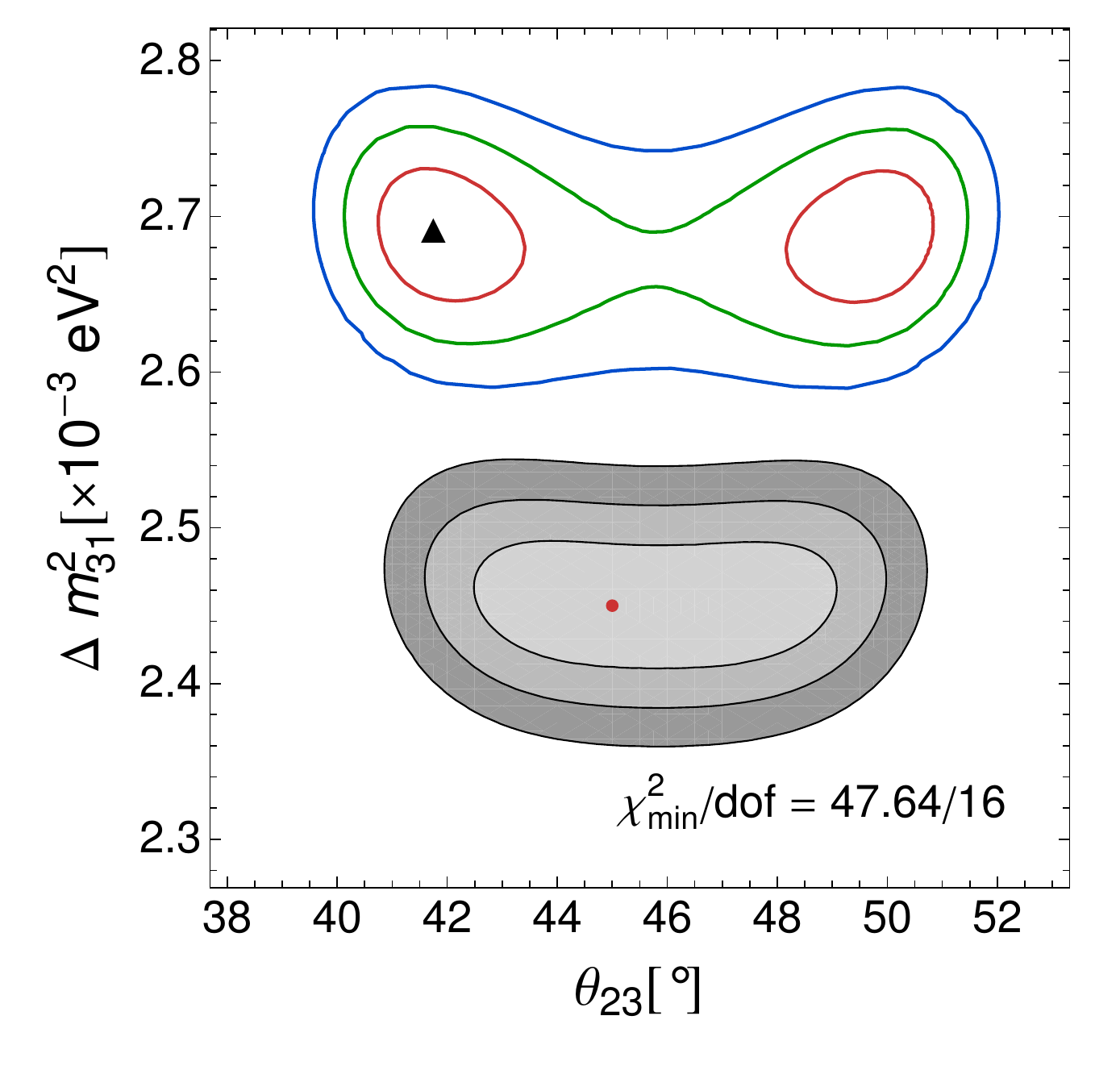}\includegraphics[width=0.5\textwidth]{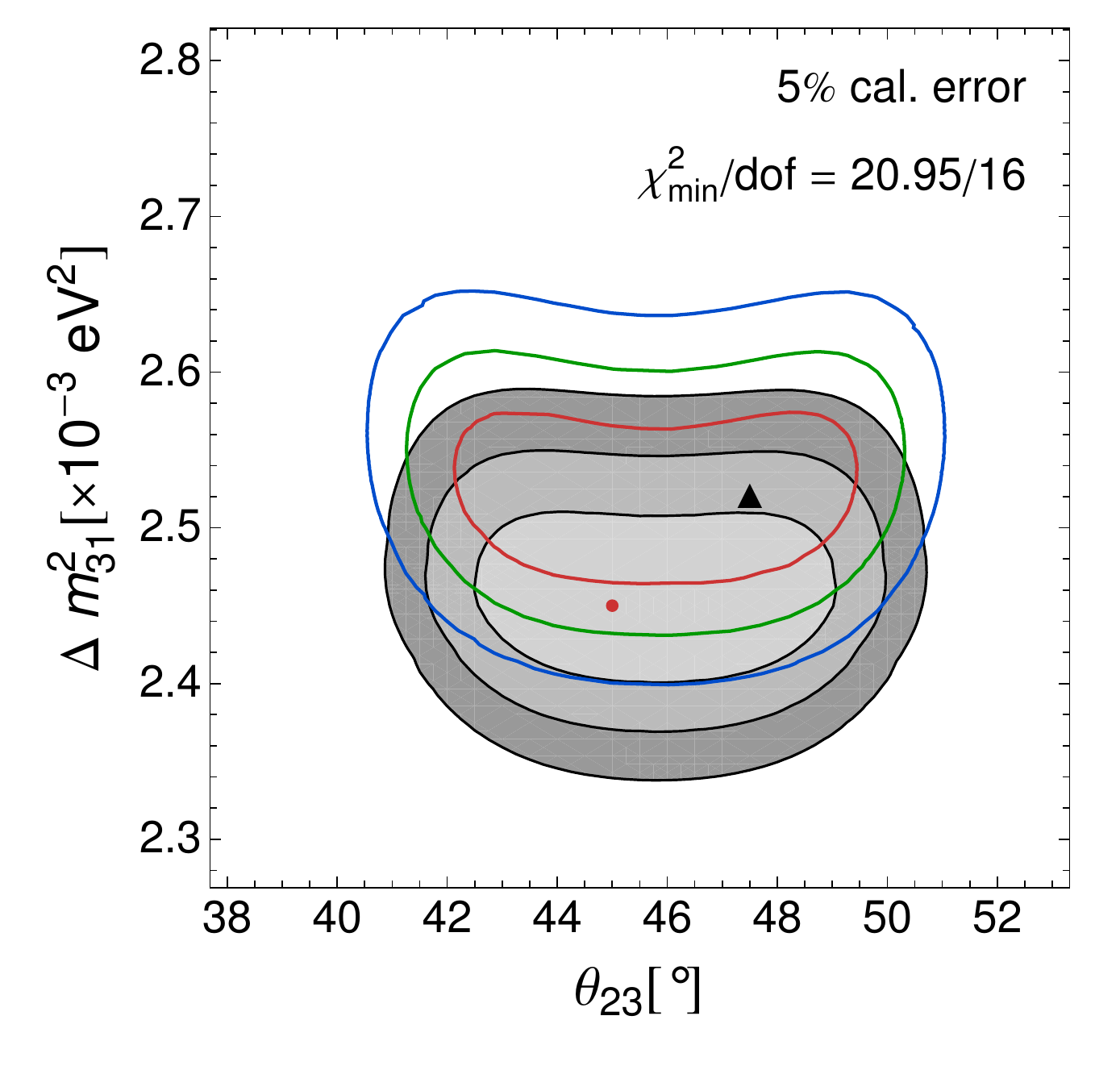}
\caption{Impact on the results if a different generator is used to
  compute the true and fitted rates in the analysis. The shaded areas
  show the confidence regions at 1, 2 and 3$\sigma$ that would be
  obtained in the $\theta_{23}-\Delta m_{31}^2$ plane if the true and
  fitted rates are generated using the same set of migration matrices
  (obtained from GiBUU, with oxygen as the target nucleus). The
  colored lines show the same confidence regions if the true rates are
  generated using matrices produced with GiBUU, but the fitted rates
  are computed using matrices produced with GENIE. Both sets of
  matrices are generated using oxygen as the target nucleus. The red
  dot indicates the true input value, while the black triangle shows
  the location of the best fit point. The value of the $\chi^2$ at the
  best fit is also shown, together with the number of degrees of
  freedom. In the left panel no energy scale uncertainty is
  considered, while for the right panel an energy scale uncertainty of
  $5\%$ is assumed, see text for details. Figure and caption adapted
  from Ref.~\cite{Coloma:2013tba} \label{fig:generators}}
\end{figure}

A common technique to estimate theory errors is to compare results
obtained with different methods by different groups and to use the
spread in results as an indicator of the associated
uncertainty. Fig.~\ref{fig:generators} shows the results of such an
analysis, utilizing a T2K-like setup, of the disappearance channel
measurement of the atmospheric parameters $\theta_{23}$ and $\Delta
m^2_{31}$.  In this analysis, data is generated with GiBUU and can
then be fitted with GENIE.  In Fig.~\ref{fig:generators}, the solid
allowed region shows the case where GibUU is used both to generate and
to fit the data.  The open regions show the results where GENIE is
used for the fitting process.  The resulting bias is very
significant. The left hand figure assumes that the energy scale is
fixed.  Given the offset in energies seen in the event rate
distributions between the two generators, see Fig.~\ref{fig:qev}, the
resulting $\chi^2$ is quite poor. In the right hand panel, the energy
scale is allowed to shift by 5\% which is enforced by the near
detector data. In this case the bias is strongly reduced and the
overall $\chi^2$ becomes much better. Nonetheless, the best fit data
point is still more than 1$\,\sigma$ away from the actual value.

We can identify two distinct, but related, problems: $\nu_e/\nu_\mu$
cross section ratios in a narrow band beam at low energy, like T2K,
and the question of energy reconstruction for both water Cerenkov
detectors and liquid argon TPCs. There are a number of steps which can
be taken to improve the situation:
\begin{itemize}
\item Better theory -- there is
considerable room for improvement, in particular, closing the gap
between event generators and theory, see for instance a recent
implementation of the spectral functions for GENIE~\cite{Jen:2014aja}.
\item More electron scattering data -- any theoretical model of the
electroweak nuclear response should be able to reproduce the somewhat
simpler electromagnetic nuclear response, which can be measured in
electron-nucleus scattering. There is a recently approved experiment
at Jefferson Lab to collect electron scattering data on
argon~\cite{Benhar:2014nca}.
\item High resolution near detector -- this is a very important
  ingredient, since a fully resolved vertex can greatly reduce model
  dependencies, but the question of flavor effects and energy
  containment remain.
\item Better flux predictions -- unlikely to reach percent level accuracy with 
superbeam sources.
\end{itemize}
A good sense of what is needed and what is believed to
be achievable can be obtained from the LBNE science
document~\cite{Adams:2013qkq}. In particular Tab.~4.5 lists the various
systematic uncertainties achieved in past $\nu_\mu\rightarrow\nu_e$
appearance searches and shows plausible extrapolations to how these might
look for LBNE. This table includes all near/far cancellations and
focuses almost entirely on rate-only effects. Furthermore, certain
crucial cancellations in the table rely on the assumption of a valid three-flavor
oscillation framework.  Finally, nearly flawless hadron 
calorimetry performance is also assumed. Interestingly, even with this 
level of assumptions, the analysis barely reaches the required 1\% goal.

The issue is then to determine what experimental strategies are available to improve
our data on cross sections so that sufficient accuracy in
long-baseline experiments becomes possible. Cross section measurements
at the percent level of accuracy will ultimately require better neutrino
sources, since a cross section measurement can, at best, be only as precise
as the accuracy with which the beam flux is known. For a detailed
understanding of detector response many exclusive, differential cross
sections have to be measured as well, for instance the energy spectrum
of neutral pions produced in neutral current interaction plays a
central role in identifying $\nu_e$ charged current events. A CP
violation measurement relies to a large degree on the comparison of
neutrino and antineutrino data and therefore, any future facility to
measure cross sections has to be able to provide neutrino and
antineutrino beams at comparable levels of precision. As explained
previously, flavor effects are non-negligible at lower energies and
therefore both $\nu_\mu$ and $\nu_e$ cross sections need to be measured
with similar accuracy. This results in the following list of
requirements for a neutrino source for precision cross section studies:
\begin{itemize}
\item Sub-percent beam flux normalization
\item Very high beam flux
\item Neutrinos and antineutrinos
\item $\nu_\mu$ and $\nu_e$
\end{itemize}

The only neutrino source which can deliver all of these characteristics is a muon storage ring, such as
nuSTORM~\cite{Adey:2013pio}. nuSTORM will deliver a beam with an equal
number of muon and electron neutrinos with a beam flux known to better
than 1\%. Storage of $\mu^-$ and $\mu^+$ allows production of
CP-conjugate beams. The very high beam intensity makes it possible to
collect a sufficient number of events within a few years in a 100\,t
near detector. The number of events for the various reaction modes is given in
Tab.~\ref{tab:nustorm}. Each event sample will at least comprise
1,000,000 events and thus statistical errors will be sufficiently small
to fully exploit the very high systematic precision offered by
nuSTORM.
\begin{table}
\begin{tabular}{|cc|cc|}
\hline
\multicolumn{2}{|c|}{$\mu^+$}&\multicolumn{2}{c|}{$\mu^-$}\\
\hline
$\bar\nu_\mu$ NC&1,174,710&$\bar\nu_e$ NC& 1,002,240\\
$\nu_e$ NC&1,817,810&$\nu_\mu$ NC&2,074,930\\
$\bar\nu_\mu$ CC& 3,030,510&$\bar\nu_e$ CC&2,519,840\\
$\nu_e$ CC&5,188,050&$\nu_\mu$ CC&6,060,580\\
\hline
\multicolumn{2}{|c|}{$\pi^+$}&\multicolumn{2}{c|}{$\pi^-$}\\
\hline
$\nu_\mu$ NC&14,384,192&$\bar\nu_\mu$ NC&6,986,343\\
$\nu_\mu$ CC&41,053,300&$\bar\nu_\mu$ CC&19,939,704\\
\hline
\end{tabular}
\caption{\label{tab:nustorm} The expected event rates for $\nu_\mu$
  and $\nu_e$ for both $\mu^+$ and $\mu^-$ circulating beams and a 100
  fiducial ton detector located 50\,m from the straight. $\nu_\mu$ and
  $\bar\nu_\mu$ rates from pion decay in the nuSTORM production
  straight are also included. The exposure is $10^{21}$ POT. Table and
  caption adapted from Ref.~\cite{Adey:2013pio}.}
\end{table}

\begin{figure}
\includegraphics[width=0.5\textwidth]{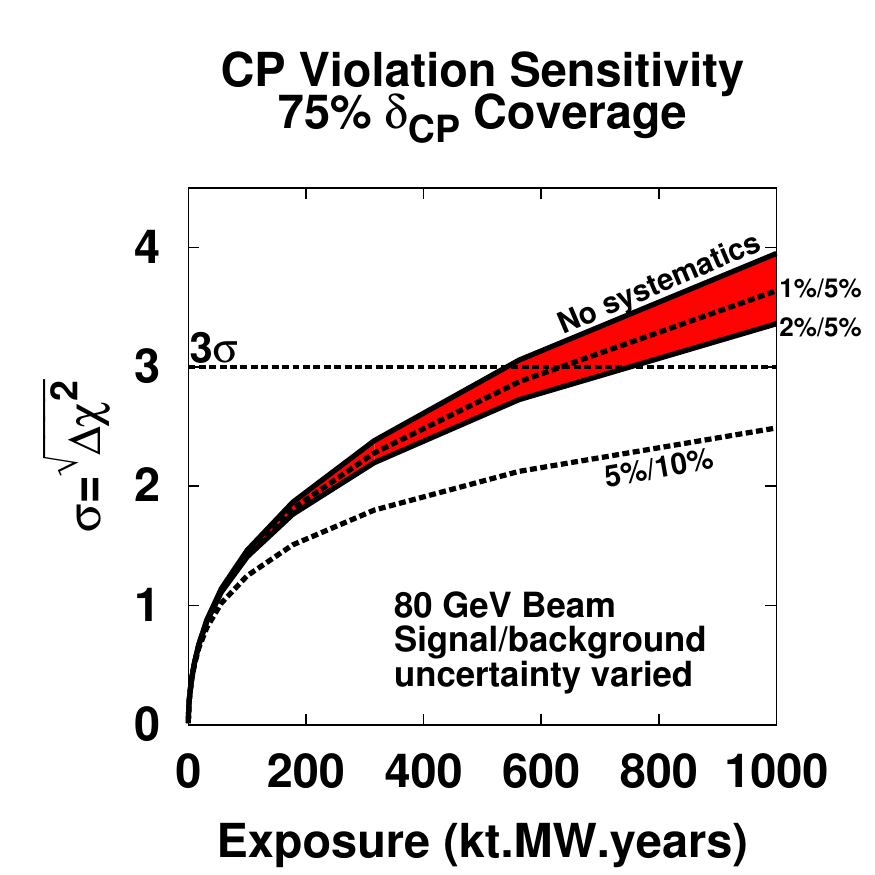}
\caption{\label{fig:lbne} Shown is the 75\% CP violation reach of LBNE
  at 3$\,\sigma$ confidence level as a function of the total exposure
  and its change under variations of the systematic error budget on
  signal normalization and background normalization
  respectively. Figure courtesy of M.~Bass.}
\end{figure}

The P5 report mandates a 3\,$\sigma$ CP violation discovery capability
over 75\% of the parameter space.  This implies that systematic
uncertainties at the 1\% level are necessary for a successful future
long-baseline program. In particular, as can be see in
Fig.~\ref{fig:lbne}, degradation of the systematic uncertainty to the
$5\%$ level corresponds to an exposure exposure increase of roughly
200-300\% in a very non-linear fashion. Current efforts to reduce
systematic errors can only credibly guarantee the upper end of the
$1-5\%$ range.  No method other than a muon storage ring such as
nuSTORM has been shown to reliably reach the lower end of this range.
Given the \$1-2 billion scale of long-baseline experiments, investing
in precise cross section measurements would provide a very good return
on investment.
\section{Summary}
\label{sec:summary}

We have made the argument that, for short-baseline physics and the
search for sterile neutrinos, a muon storage ring such as nuSTORM would 
provide a unique facility of unrivaled capability.  It would provide access to 
all possible oscillation channels involving muon and electron neutrinos and, 
in the case of a sterile neutrino discovery, would allow detailed exploration of 
the underlying physics model.

\begin{figure}
\includegraphics[width=0.6\textwidth]{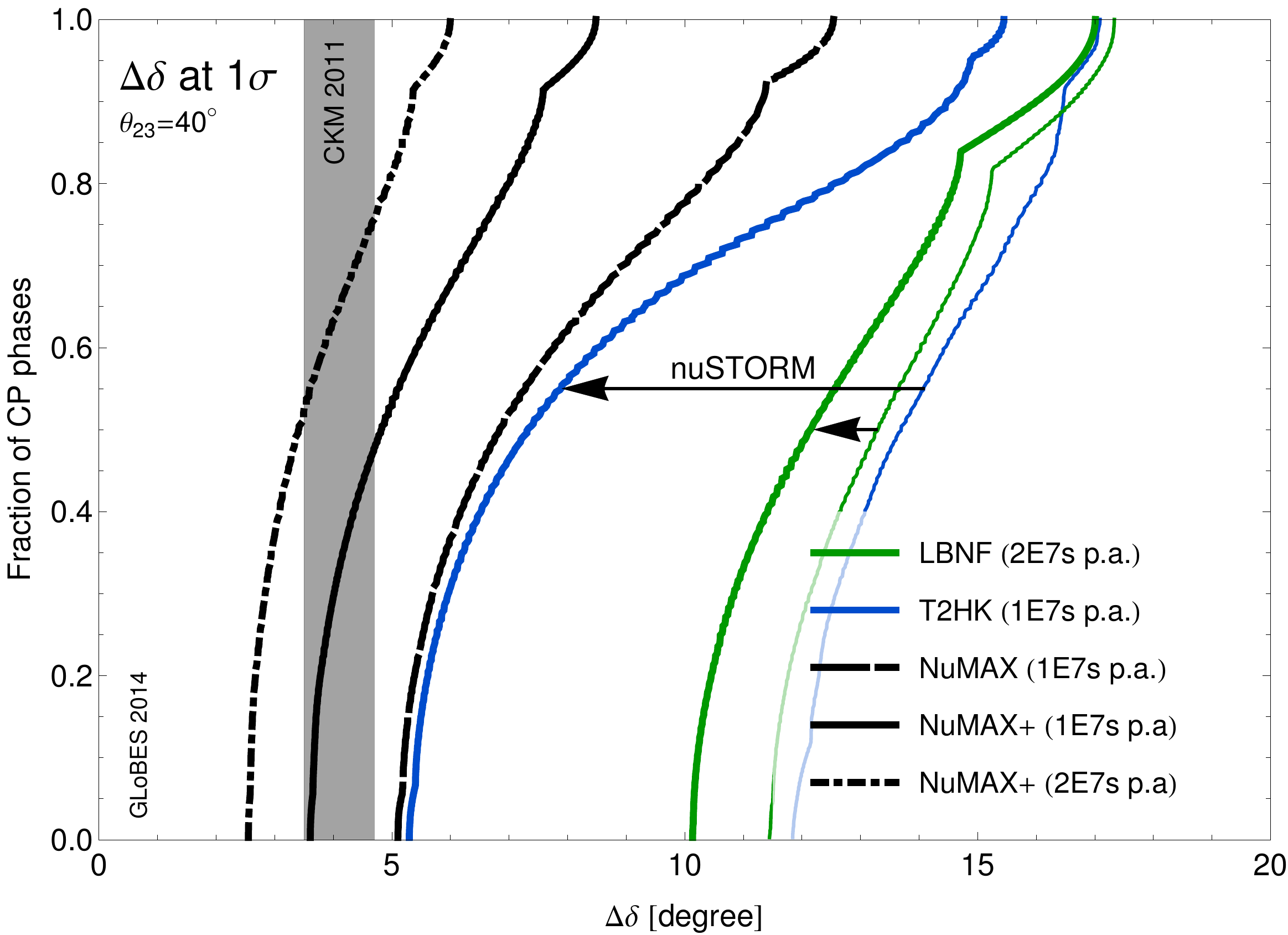}
\caption{\label{fig:bamboo} Expected precision for a measurement of
  $\delta$ at future long baseline oscillation experiments. Results
  are shown as a function of the fraction of possible values of
  $\delta$ for which a given precision (defined as half of the
  confidence interval at $1\sigma$, for 1 d.o.f.) is expected. All
  oscillation parameters are set to their present best fit values, and
  marginalization is performed within their allowed intervals at
  $1\sigma$, with the exception of $\theta_{13}$ for which
  marginalization is done within the allowed interval expected at the
  end of the Daya Bay run. Matter density is set to the value given by
  the PREM profile, and a 2\% uncertainty is considered. The hierarchy
  is assumed to be normal, and no sign degeneracies are accounted
  for. Systematic uncertainties are implemented as in
  Ref.~\cite{Coloma:2012ji}. All facilities include an ideal near
  detector, and systematics are set to their `default' values from
  Tab.~2 in Ref.~\cite{Coloma:2012ji}.  The nominal running time is 
  10 years for each experiment.  The different lines correspond
  to the following configurations.  \textbf{LBNF} corresponds to the P5 endorsed
  version of the restructured LBNE project. The LBNE CDR~\cite{CDR} beam flux has been
  used. The detector performance has been simulated as in
  Ref.~\cite{CDR}, using migration matrices for NC backgrounds
  from Ref.~\cite{Akiri:2011dv}. A 40\,kton detector and 1.2\,MW beam
  power are assumed.  \textbf{T2HK} stands for a 750 kW beam directed 
  from Tokai to the Hyper-Kamiokande detector (560 kton fiducial mass)
  in Japan. The baseline and off-axis angle are the same as for
  T2K. The detector performance has been simulated as in
  Ref.~\cite{Coloma:2012ji}.  \textbf{nuSTORM}, the curves with an
  arrow pointing to them, assume that nuSTORM has delivered a 1\%
  measurement of the relevant cross sections. \textbf{NuMAX}
  corresponds to a low-luminosity neutrino factory obtained from the
  decay of 5 GeV muons, simulated as in
  Ref.~\cite{Christensen:2013va}. The beam luminosity is set to
  $1.9\times10^{20}$ useful muon decays of each polarity per $10^7$\,s,
  and the flux is aimed to a 40\,kton magnetized LAr detector placed
  at 1300 km from the source. \textbf{NuMAX+} corresponds to
  $5\times10^{20}$ useful muon decays of each polarity per $10^7$\,s. All
  calculations are performed with
  GLoBES~\cite{Huber:2004ka,Huber:2007ji}.}
\end{figure}

Furthermore, in the context of the future long-baseline program, we discussed 
the need for an improved understanding of neutrino-nucleon cross sections. 
In order to achieve the P5 goal of 3\,$\sigma$ discovery potential over 75\% 
of all CP phases, a stringent systematic uncertainty goal of $\le1\%$ must be met. 
Currently, no method other than 
the precision neutrino beams from a muon storage ring can guarantee this level of 
accuracy with a high degree of certainty. The impact of a precise cross section 
determination on various long-baseline experiments and their ability
to measure the CP phase $\delta$ is shown in Fig.~\ref{fig:bamboo}.

It has been recognized for many years that a neutrino factory will be
the ultimate tool in exploring neutrino oscillations and this
statement remains true even when $\theta_{13}$ is large. With a view to
the long-term evolution of the global and domestic neutrino programs, a
staging scenario was developed~\cite{Delahaye:2013jla}, which exploits the unique facilities
which will be created for LBNF in the best possible way. This leads to
the NuMAX concept~\cite{Christensen:2013va} and corresponding
sensitivities are shown in Fig.~\ref{fig:bamboo}. A neutrino factory,
like NuMAX+ would enable CP violation to be measured in the neutrino 
sector with the same accuracy as has been achieved in the quark sector.
%

Muon-based neutrino beams offer the precision measurement capabilities required to ensure 
success in meeting the neutrino-related science goals as outlined by P5.  In the near- to mid-term, 
a muon storage ring, such as nuSTORM, would provide the capabilities required to mitigate the 
\emph{otherwise substantial risk} that LBNF might not achieve its sensitivity goal, as set by P5, for 
discovering CP violation in the neutrino sector.  Furthermore, such a ring would enable a truly 
definitive search for sterile neutrinos.  In the longer term, beyond LBNF, more advanced muon 
accelerator capabilities, such as NuMAX, would provide the tools required for precision studies of 
CP violation in the neutrino sector:
\begin{itemize}
\item NuMAX -- precision CP phase
\item NuMAX+ -- high precision CP phase and unitarity
\end{itemize}
This leads to the conclusion that a high priority should be placed on maintaining the research 
effort towards muon accelerator capabilities as part of the ongoing accelerator R\&D portfolio.

\acknowledgments

 This work has been supported by
the U.S.  Department of Energy under award number
\protect{DE-SC0003915} and contract \protect{DE-AC02-07CH11359}.

\bibliographystyle{apsrev} 
\bibliography{references}

\end{document}